\documentclass{imsart}

\usepackage{natbib}
\usepackage{comment}

\usepackage{amsmath,bm,amsfonts,color,amsthm}
\usepackage{graphicx}

\usepackage{caption}
\usepackage{subcaption}

\usepackage{float}
\restylefloat{table}

\newcommand{\commentt}[1]{}

\newcommand{\bx}{\boldsymbol{x}}

\newcommand{\by}{\boldsymbol{y}}

\usepackage{mathtools}	

\usepackage[switch]{lineno}
\RequirePackage[colorlinks,citecolor=blue,urlcolor=blue]{hyperref}
\usepackage[ruled,lined]{algorithm2e}
\SetKw{KwSet}{Set}

\begin{document}
\begin{frontmatter}
\title{ Blocking Methods Applied to Casualty Records from the Syrian Conflict}
\runtitle{Blocking Methods Applied to Syrian Conflict}
\begin{aug}
  \author{\fnms{Peter} \snm{Sadosky}\ead[label=e1]{psadosky@gmail.com}}, 
    \author{\fnms{Anshumali} \snm{Shrivastava}\ead[label=e2]{anshumali@rice.edu}}, \\
    \author{\fnms{Megan} \snm{Price}\ead[label=e3]{meganp@hrdag.org}}, 
    \and
  \author{\fnms{Rebecca C.} \snm{Steorts}\ead[label=e4]{beka@stat.duke.edu}} 

  \runauthor{Sadosky et. al}
  \affiliation{Carnegie Mellon University\\ Undergraduate Student}
  \address{Carnegie Mellon University\\ Economics, History, and Statistics Departments\\ Undergraduate Student\\
  5000 Forbes Avenue, Pittsburgh, PA 
    15213\\ \printead{e1}}
     \address{Rice University\\Computer Science Department\\ Assistant Professor\\
    Houston, Texas 77005\\ \printead{e2} } 
     \address{Human Rights Data Analysis Group (HRDAG)\\ Director of Research\\
   San Francisco, CA 94110, USA\\ \printead{e3}}
   \address{ Duke University\\ Department of Statistical Science\\  Assistant Professor\\
   Durham, NC 27708, USA\\ \printead{e4}}
  \end{aug}


\begin{abstract}
Estimation of death counts and associated standard errors is of great importance in armed conflict such as the ongoing violence in Syria, as well as historical conflicts in Guatemala, Per\'u, Colombia, Timor Leste, and Kosovo. For example, statistical estimates of death counts were cited as important evidence in the trial of General Efra\'in R\'ios Montt for acts of genocide in Guatemala. Estimation relies on both record linkage and multiple systems estimation.  A key first step in this process is identifying ways to partition the records such that they are computationally manageable.  This step is referred to as blocking and is a major challenge for the Syrian database since it is sparse in the number of duplicate records and feature poor in its attributes. As a consequence, we propose locality sensitive hashing (LSH) methods to overcome these challenges. We demonstrate the computational superiority and error rates of these methods by comparing our proposed approach with others in the literature. We conclude with a discussion of many challenges of merging LSH with record linkage to achieve an estimate of the number of uniquely documented deaths in the Syrian conflict. 


\end{abstract}


\end{frontmatter}

\section{Introduction}
The Syrian conflict has been at the center of international news  since March 2011. One main question is ``how many civilian casualties have occurred as a result of this conflict?" Although this conflict is well documented, answering what might seem to be a simple question is statistically challenging. This stems from the fact that some victims may be reported to multiple sources whereas other victims are not reported at all. We are motivated by finding a death estimate (with associated standard errors) as such information may contribute to future transitional justice and accountability mechanisms. For instance, statistical estimates of death counts have been introduced as evidence in national court cases and international tribunals investigating the responsibility of state leaders for crimes against humanity. 

Reporting an estimate is a multi-step process, where the first involves data reduction via a process referred to as blocking. Blocking partitions the space of records into similar ``blocks" or groups. The second step involves record linkage (within each block), which is the process of merging many noisy databases to remove duplicate entities. On any moderately sized database it is essential to avoid all-to-all record comparisons, thus emphasizing the importance of the initial blocking step and the use of algorithms that are computationally fast. The third step involves multiple systems estimation, where post-blocking and post-record linkage, we attempt to estimate the total number of entries in a closed population. All components of this multi-step process are challenging and complex, and thus, this paper 
 focuses on advancements in the first step, blocking, for the Syrian conflict. 

We illustrate that traditional blocking methods typically split records referring to the same person across different blocks around 50\% of the time, which makes the record linkage step pointless. However, we propose combining the work of  \cite{shrivastava2014densifying,shrivastava2014improved} and applying these approaches to the Syrian database. In this combined approach, we find that this method  only splits records referring to the same person across different blocks less than 1\% of the time. Furthermore, this method is linear in the tuning parameters. We illustrate using our \texttt{Java} package that for the approximately 300,000 death records, our blocking procedure runs in 10 minutes. See \S \ref{sec:disc} for a discussion of the remaining challenges of integrating our proposed methods for blocking based on \cite{shrivastava2014densifying,shrivastava2014improved} with any subsequent record linkage procedure.

\subsection{Prior Work}
\label{sec:priorwork}
Record linkage, also known as coreference resolution, entity resolution, and de-duplication
is a well known but difficult problem, especially for real world applications, which include human rights violations, official statistics, medical applications, and others
\citep{christen_2011, Herzog_2007,Herzog:2010}.  
Such obstacles are due to the noise inherent in the data, which is often hard to
accurately model \citep{pasula_2003, steorts_2015_jasa}.  A
more substantial obstacle is the scalability of the approaches \citep{WYP:2010}.  
Assuming $d$ databases of $n$ records each, brute-force approaches,
using all-to-all comparisons, require $O(n^d)$ comparisons.  Such approaches are easily
prohibitive for moderate $n$ or $d$.  To avoid this computational
burden, the number of comparisons made must be drastically reduced, without
compromising linkage accuracy.  The record linkage literature tries to
achieve scalability by blocking, which involves partitioning records into
``blocks'' and treating records in different blocks as non-co-referent {\em a
  priori} \citep{christen_2011, Herzog_2007}. Record linkage methods are only
applied {\em within} blocks, reducing the comparisons to
$O(B n_{\max}^d)$, with $n_{\max}$ being the size of the largest of the $B$
blocks. 

There are several techniques  for constructing a blocking partition.
The most basic method picks certain fields (e.g., governorate, or sex and year
of death) and places records in the same block if and only if they agree on
all such fields.  This amounts to an {\em a priori} judgment that these fields
are error-free. We refer to this as \emph{traditional blocking} (see \S \ref{sec:blocking}).

Other data-dependent blocking methods \citep{christen_2011, WYP:2010}
 are highly application-specific or are based on placing similar records into the
same block, using techniques of random projections or locality sensitive hashing (LSH)~\citep{indyk_1998}. 
LSH is a \emph{probabilistic method} of dimension
reduction, which is widely used in computer science and in database
engineering as a way of rapidly finding approximate nearest neighbors
\citep{gionis_1999}.  
  Unlike conventional blocking, LSH uses all
the fields of a record, and can be adjusted to ensure that blocks are
manageably small, but then do not allow for further record linkage within blocks.  
Such methods are fast and have high recall (true positive rate), but suffer from low precision, rather, too
many false positives.

\cite{steorts_2014_hash} proposed clustering-based blocking schemes that are variants on LSH.  The first, transitive locality sensitive hashing (TLSH) is based upon the community discovery literature such that \emph{a soft transitivity} (or relaxed form of transitivity) can be imposed across blocks. The second, $k$-means locality sensitive hashing (KLSH) is based upon the information retrieval literature and  clusters similar records into blocks using a vector-space representation and projections (KLSH had been
used before in information retrieval but never with record linkage
\citep{pauleve_2010}). \cite{steorts_2014_hash} showed that both KLSH and TLSH gave improvements over popular methods in the literature such as traditional blocking, canopies \citep{mccallum_2000}, and $k$-nearest neighbors clustering. 

There are many variants of LSH and one popular form is minwise hashing. All LSH methods are defined by a type of similarity and a type of dimension reduction \citep{broder_1997}.  
Recently, \cite{shrivastava2014densifying} showed that minwise hashing based approaches are superior to random projection based approaches when the data is very sparse and feature poor. Furthermore, 
improvements in computational speed can be obtained by using the recently proposed densification scheme known as densified one permutation hashing (DOPH)~\citep{shrivastava2014densifying,shrivastava2014improved}.  Specifically, the authors proposed an efficient substitute for minwise hashing, which only requires one permutation (or one hash function) for generating many different hash values needed for indexing. In short, the algorithm is linear (or constant) in the tuning parameters, making it very computationally efficient. 

\S \ref{sec:motivate} provides a brief background on the Syrian conflict, as well as a description and challenges of the data. \S \ref{sec:blocking} reviews the blocking literature that has been explored for records from the Syrian conflict. \S \ref{hashing} introduces an ensemble of hashing methods that we propose as an application for blocking the Syrian database. \S \ref{sec:app} applies our proposed method for hashing to the Syrian database, illustrating that we only split records across blocks less than 1\% of the time, compared to at best 20\% of the time for every other blocking method that was considered.  
We give a thorough discussion of the challenges of integrating LSH with record linkage for the Syrian database in \S \ref{sec:disc}.

\section{Motivation: The Syrian Conflict}
\label{sec:motivate}
Violence broke out in Syria in March 2011 following a series of anti-government protests.  In the years since then the conflict has continued to escalate.  Restrictions on both local and international media make it difficult to determine the scope of the violence \citep{FH2012, CJR2014}.  Nonetheless, combinations of conventional and citizen journalists, grassroots non-governmental organizations (NGOs), and international humanitarian organizations continue to do their best to document the conflict.  For example, early in the conflict, Human Rights Watch (HRW) documented numerous cases of lethal force used against peaceful protestors, and confirmed from defectors from Syria's security forces that ``\ldots they were given orders to fire on unarmed protesters.''   HRW further reported that ``\ldots other aspects of the repression - arbitrary and incommunicado detention, rampant torture, and denial of medical care –have continued unabated" \citep{HRW2011}.  More recently, as summarized by the BBC, the United Nations Commission of Inquiry ``\ldots has evidence that those on both sides of the conflict have committed war crimes - including murder, torture, rape and enforced disappearances. Government and rebel forces have also been accused by investigators of using civilian suffering, such as blocking access to food, water and health services, as a method (of) war.''  Chaos in the region has also been credited with creating an opportunity for the Islamic State to take control of large portions of territory.  Fighting now involves a large number of armed groups with shifting allegiances and boundaries.  Numerous attempts by members of the international community to broker ceasefires and negotiate dialogues have thus far been unsuccessful \citep{bbc2015}.

Our motivation is to estimate the number of conflict related killings in Syria since March 2011. Such estimation is necessary since many acts of violence are hidden and many victims may not be reported or identified until months or even years after the event.  Relying solely on what is observable is inadequate.  As described by Maria McFarland, Co-Director of the US Program for Human Rights Watch, in her testimony to the Tom Lantos Human Rights Commission, ``\ldots collecting information about what is happening in Syria today is extremely difficult. The situation we have been able to document is extremely disturbing, but we are just as concerned by what we do not know and have been unable to confirm as by what we do know'' \citep{HRW2012}.

Machine learning and statistical modeling methods can be applied to information about documented, identifiable victims to estimate a total number of victims, both those currently identified and those not yet identified or documented.  These efforts may contribute to future accountability and transitional justice mechanisms in Syria.  Additionally, methodological developments for this particular application may support similar efforts following other violent conflicts. A key first step in this analysis is identifying computationally tractable blocking methods to enable record linkage and further statistical modeling. 
 
\subsection{The Data}

Via collaboration with the Human Rights Data Analysis Group (HRDAG), we have access to four databases. They come from the Violation Documentation Centre (VDC), Syrian Center for Statistics and Research (CSR-SY), Syrian Network for Human Rights (SNHR), and Syria Shuhada website (SS). Each database lists each victim killed in the Syrian conflict, along with identifying information about each person (see \cite{price_2013} for further details).

Data collection by these organizations is carried out in a variety of ways. Three of the groups (VDC, CSR-SY, and SNHR) have trusted networks on the ground in Syria.  These networks collect as much information as possible about the victims. For example, information is collected 
through direct community contacts. Sometimes information comes from a victim's friends or family members. Other times, information comes from religious leaders, hospital, or morgue records.  These networks also verify information collected via social and traditional media sources.  The fourth source, SS, aggregates records from multiple other sources, including NGOs as well as social and traditional media sources (see \url{http://syrianshuhada.com/} for information about specific sources). 

These lists, despite being products of extremely careful, systematic data collection, are not probabilistic samples \citep{Sig2015, IOAS2015, CJLS2015, price2014updated}. Thus, these lists cannot be assumed to represent the underlying population of all victims of conflict violence.  Records collected by each source are subject to biases, stemming from a number of potential causes, including a group's relationship within a community, resource availability, and the current security situation.  Although it is beyond the scope of this paper, final analyses of these sources must appropriately adjust for such biases before drawing conclusions about patterns of violence.

As already mentioned, our ultimate goal is to merge the four databases, each having a different number of recorded victims, so as to remove duplicate entities among them. In this respect, before being able to use record linkage or multiple systems estimation, we first must use blocking to reduce the space from all-to-all record comparisons. To help assess any blocking method, we have a set of training data generated through hand-matching. Four different matchers from HRDAG manually reviewed records in the databases, classifying records that referred to the same individual as matches, and records with no possibility of matching as non-matches. We treat the hand-matched data as a gold standard against which to compare our blocking approaches.


\subsection{Unique Challenges Posed by Data from Syrian Conflict}
With roughly 300,000 total death records, the total possible pairs of record comparisons are of the order of $10^{10}$. The number of comparisons to be made must be drastically reduced,
without compromising linkage accuracy. Hence, we turn to blocking, or rather, placing similar records in groups 
or partitions. 

There are two main features of the data that make traditional blocking approaches difficult, if not impossible. The first is due to the fact that the data is very sparse in the number of duplicates, and the second is that the data is sparse in the features that can be used to reduce the number of all-to-all record comparisons. In the following section, we introduce traditional blocking methods for record linkage, and then show that such blocking methods perform so poorly that record linkage would never be performed. In particular, note that for the Syrian database, the traditional or simplistic forms of blocking induce errors in the sense that records that refer to the same person are often placed into separate blocks, which then cannot be recovered from any record linkage algorithm. This makes traditional types of blocking inherently impossible on the Syrian database. This calls for special blocking methods to be used on \emph{sparse, feature-poor} data.

\section{Blocking}
\label{sec:blocking}
Blocking is a set of rules or algorithms that reduces the set of all-to-all record comparisons. Blocking divides records into mutually exclusive and jointly exhaustive blocks or partitions,
allowing record linkage to be performed within each block \citep{winkler_2006, steorts_2014_hash}. Only records
within the same block can be linked and linkage algorithms aggregate
information across blocks. The most basic method for constructing a blocking partition picks certain
features (e.g. same last name, date of death (DoD), place of death, etc.) and places records in the
same block if and only if they agree on all such fields. This amounts to an
a priori judgment that these fields are error-free. We call this traditional
blocking as was first coined in \cite{steorts_2014_hash}. In the setting of the Syrian conflict, 
traditional blocking is not realistic since many blocks are so large that linkage is computationally intractable. Also, since blocks only consider selected features, much time
is wasted comparing records that happen to agree on these features but clearly refer to different individuals. 
Since traditional blocking proves difficult, we next review conjunctions, conjunctions combined with Arabic soundex, and random projections. 

\subsection{Conjuctions}
\label{sec:conjuctions}
A conjunction links two sets of variables through the operators union $\cup$ and intersection $\cap$ \citep{michelson:aaai06}. 
Suppose we have two variable fields, date of death and governorate. 
Define a sample conjunction as $DoD \cap Governorate$, which creates a partition for each set of records that agree in these two fields.
If two records have a date of death of 2013-06-20 and are from Damascus, then they will be placed in the same partition. A third record, with the same date of death but from Homs, will not meet the rule and will be placed in a second partition.
A conjunction  specifies the relationship that must exist for two records to be joined together.
As a result, each partition contains records that are in some way similar, or agree in a subset of their features.

A disjunction of conjunctions is simply the union of two or more conjunction rules. 
An example of a disjunction of conjunctions is $[(A \cup B) \cap (B \cup C)]$ for a set of record fields $A, B, C, D$. 
As the number of rules grows, a higher degree of similarity between two records is required to form a candidate pair.

A disjunction of conjunctions approach is used on the Syrian dataset as a way of partitioning records by their features. 
The input to a conjunction rule, in this context, is a feature from the records, and the output is a set of candidate pairs to be compared to ground truth data. 
The main benefits of using conjunctions for blocking records are the ability to target an approximate number of candidate pairs and ease in isolating certain notable features about the data. For example, if we see a few large clusters of reported locations, then we could subset those records easily. 
The main disadvantages of using conjunctions is the possibility of over- or under-fitting the data, scalability to moderate/large data, and each conjunction is application specific. 
Thus, the main challenge is in constructing an optimal set of rules for the conjunction scheme. 
For $m$ variable fields, there are a minimum of $m^m$ possible combinations of conjunctions, with this number growing as variable fields are subsetted (as in subsetting date of death into its year, month, and day components).

\subsection{Conjunctions and Arabic Soundex}

Often, a conjunction scheme by itself is not enough to partition the records effectively. 
This is due to attributes of the data, including incomplete records, typos, and feature-poor data.
In this case, we propose combining a disjunction of conjunctions approach with textual string analysis of the provided Arabic field names. (See \cite{price_2013} for a description of previous considerations of textual string analysis of Arabic names, in particular comparing pairs in which one name is recorded in English and the other is recorded in Arabic).
This approach allows us to make use of the name variable in the data to further segment and partition records by similarity.  We create partitions as small and accurate as possible, where candidate pairs are generated only if they match with the ground truth data. 

Here, we incorporate the strongest conjunction blocks from the \S \ref{sec:conjuctions} with an Arabic Edit Distance Algorithm (AEDA) \citep{abdel_2011}. 
The AEDA algorithm is intended to be an extension of a Levenshtein distance measure for Arabic text. 
The Levenshtein distance is a metric for comparing the similarity of two strings; it works by finding the minimum number of character edits (either through insertion, deletion, or substitution) needed to convert one string into another. 
Arabic text does not operate in this way, as \cite{abdel_2011} suggests, since Arabic words can often be almost identical in characters but have a very low degree of similarity in meaning or intention. 
Thus, we use AEDA as a way of finding similarity between two Arabic strings, and not just checking for exact matches.
Once we can determine how similar two name fields are, we can determine whether two records should remain in the same partition or be separated. 

Let $a,b$ be any two Arabic characters, and using AEDA we wish to know the cost associated with converting $a$ into $b.$
If the characters are very similar, then the cost to replace them should be very low.
As they become more dissimilar, the cost to replace the first character with the second should increase.   
There are three possible replacement costs to consider in this case: phonetic replacement $\alpha(a, b)$, letter form replacement $\beta(a, b)$, and keyboard distance replacement $\gamma(a, b)$. 
Each cost is a function of the two input characters.
If $a = b$, then the replacement cost for each of the three functions is 0. 
Let
$\omega, \lambda, \sigma$ be the weights associated with each of the costs, respectively.  
The AEDA formula (Equation~\ref{eqn:fry}) incorporates these three functions into a single formula, where we must decide how much weight should be given to each feature. 
\begin{align}
 \text{frc}(a, b) = \begin{dcases*}
        \frac{\alpha(a, b)\cdot\omega + \beta(a, b)\cdot\lambda+\gamma(a, b)\cdot\sigma}{\omega+\lambda+\sigma}  & if a $\neq$ b\\
        0 & Otherwise
        \end{dcases*}  
  \label{eqn:fry}       
\end{align}

For determining phonetic similarity, the first cost, the two provided Arabic characters are evaluated for how they sound to a native Arabic speaker, with values for pairwise characters on a scale from 0 to 1. 
\cite{abdel_2011} provides a table with these scores. 
A score of 0 on this scale means that the characters have no similarity, while a 1 means that they are the same letter or sound very similar. 
For letter form replacement, characters that are often swapped or mistakenly used in place of one another by an Arabic writer define a measure of similarity. 
This form is especially useful in transcription issues, where two characters could be used for the same spoken sound. 
The third cost is a keyboard distance, which indicates that characters close to each other on an Arabic keyboard are more similar to one another than characters far away, as they are more likely to have been accidentally transcribed.

These three functions attempt to capture the possible transcription issues present in tracking and recording casualty counts for a database.
Depending on what methods are used to record and transcribe the information provided, one replacement cost may be more necessary than another.

The keyboard relationship is given by $\text{Sim}_{Kb}(a,b) = 1 - \frac{\sqrt{(x_{a}-x_{b})^{2}+(y_{a}-y_{b})^{2}}}{\psi},$ where
 ${Sim}_{Kb}$ is the keyboard similarity between two characters $a$ and $b$. 
The similarity of the two characters is found by calculating the x and y distances between the two characters on the keyboard, where $\psi$ is the maximum distance between any two characters on the keyboard (which is 12 units on standard Arabic keyboards).
The x and y distances can be found by looking at an Arabic keyboard (or looking at a picture online), and noting the horizontal and vertical distance between the two input characters.

Once these three similarity measures are computed individually, the replacement cost $frc$ formula (Equation~\ref{eqn:fry}) is used to balance the weighting/importance of each. 
The $\omega$, $\lambda$, and $\sigma$ are the weights given to each cost function, all summing to 1, and must be chosen with care and with respect to the transcription problems identified in the data. 
For example, if all of the possible error between two strings is in an auditory transcription from a speaker to a typist, then the weighting should all go towards the phonetic component.
If no information is known about the data recording process, providing an equal weighting to each term is likely the best option.

Once records have been blocked based on conjunction rules, they can be further partitioned based on the results of the replacement cost formula. 
Pairwise records with low cost values from this function are partitioned and generate final candidate pairs, while records with high cost values are no longer considered as possible pairs.
This second partitioning reduces the space of comparisons needed to be made to the ground truth data, which in some cases can dramatically reduce the computational time required for the method to run.

%
%


\section{Ensemble of Hashing Methods}
\label{hashing}
We consider an ensemble of hashing methods, which we use for comparison on the Syrian database (\S \ref{sec:app}). They are all based upon LSH, namely, KLSH,
minhashing and weighted minhashing. 
LSH-based blocking schemes ``shingle''
\citep{rajaraman_2012} records.  That is, each record is treated as a string and
is replaced by a ``bag'' (or ``multi-set'') of length-$k$ contiguous
sub-strings that it contains. These are known as ``$k$-grams'', ``shingles'',
or ``tokens''.  The string ``TORONTO'' yields the bag of length-two
shingles ``TO'', ``OR'', ``RO'', ``ON'', ``NT'', ``TO''.  (N.B., ``TO'' appears
twice.)
As an alternative to shingling, we might use a bag-of-words (BoW) representation, or
even to shingle into consecutive pairs (triples, etc.) of words. 
We first describe some basics of hashing, and then describe KLSH, and then minhashing and weighted minhashing. Minhashing and weighted minhashing are sped up using DOPH (see \S \ref{sec:doph}.)

In LSH, a hash function is defined as $y = h(x),$ where $y$ is the \emph{hash code} and $h(\cdot)$ the \emph{hash function}. A \emph{hash table} is a data structure that is composed of buckets (not to be confused with blocks), each of which is indexed by a hash code. Each reference item $x$ is placed into a bucket $h(x).$ For a review of LSH, we refer to \cite{rajaraman_2012}. 

\subsection{KLSH}
We explore a simple random projection method, $k$-means locality sensitive hashing (KLSH). In KLSH,  the number of times each shingle type appears in a record is counted, leading to a bag-of-shingles representation
for records. For measuring similarity between records, the inner
product of bag-of-shingled vectors (of records) is used, with inverse-document-frequency weighting. Then the bag-of-shingled vectors is reduced first using random projections and second by clustering the low-dimensional projected
vectors via the $k$-means algorithm. To put it simply, the mean number of records per cluster is controlled by $n/c,$ where $n$ is the total number of records and $c$ is the number of block-clusters \citep{steorts_2014_hash}.

\subsection{Minhashing}
One of the most popular forms of LSH is known as minhashing, where the similarity between records is Jaccard \citep{Proc:Broder_STOC98}.  Let $\{0,1\}^D$ denote the set of all binary $D$ dimensional vectors, while $\mathbb{R}^D$ refers to the set of all $D$ dimensional vectors (of records). 
The records can be represented in vector representation via shingling or a BoW method. 
Given two sets (or equivalently binary vectors) $x,y \in \{0,1\}^D,$
the Jaccard similarity between $x, y \in \{0,1\}^D$  is\begin{align*}
\mathcal{J} = \frac{|x \cap y|}{|x \cup y|},
\end{align*}
where $|\cdot|$ is the cardinality of the set.  Since we use a shingling based approach, our representation of each record is likely to be very sparse. Moreover, \cite{shrivastava2014defense} showed that minhashing based approaches are superior than the random projection based approaches for very sparse datasets.

The minwise hashing family applies a random permutation $\pi$, on the given set $S$, and stores only the minimum value after the permutation mapping, known as the \emph{minhash}.  Formally, the minhash is defined as $h_{\pi}^{min}(S) = \min(\pi(S))$, where $h(\cdot)$ is a hash function.

Given sets $S_1$ and $S_2$, it can be shown by an elementary probability argument that
\begin{equation}
\label{eq:MinHash}
Pr_{\pi}({h_{\pi}^{min}(S_1) = h_{\pi}^{min}(S_2)) =  \frac{|S_1 \cap S_2|}{| S_1 \cup S_2|}},
\end{equation}
where the probability is over uniform sampling of $\pi$. It follows from Equation~\ref{eq:MinHash} that minhashing is an LSH for Jaccard similarity.


\subsection{Making Minwise Hashing practical: Densified One Permutation Hashing (DOPH)}
Let $K$ be the number of hash functions and let $L$ be the number of hash tables. 
A  $(K,L)$ parameterized blocking scheme requires $K \times L$ hash computations per record. For a single record, this requires storing and processing hundreds (or even thousands) of very large permutations.  This in turn requires hundreds or thousands of passes over each record. Thus, traditional minwise hashing is prohibitively expensive for large or moderately sized datasets. In order to cross-validate the optimal $(K,L)$ tuning parameters, we need multiple independent runs of the $(K,L)$ parameterized blocking scheme. This expensive computation is a major computational concern.

Instead we utilize the methods of \cite{shrivastava2014densifying,shrivastava2014improved}, which use only \emph{one permutation} and computes $k = K \times L$ minhashes with the required property (Equation~\ref{eq:MinHash}) in just one pass over the data.  Furthermore, 
due to sparsity of data vectors (from shingling), empty buckets (in the hash tables) are possible and destroy LSH's essential property \citep{rajaraman_2012}.
To restore this, 
we rotate the values of non-empty buckets and assign a number to each of the empty buckets. 
Our $KL$ hashed values are simply the final assigned values in each of the $KL$ buckets. The final values were shown to satisfy Equation~\ref{eq:MinHash}, for any $S_1, \ S_2$, as in minhash~\citep{shrivastava2014densifying,shrivastava2014improved}.

\subsection{Weighted Densified One Permutation Hashing}
\label{sec:doph}

Minhashing, however, only uses the binary information and ignores the weights (or values) of the components, which as argued before are important for the problem due to the sparsity and feature-poor data (see \S \ref{sec:priorwork}). This is the reason why we observe slightly better performance for synthetic data of LSH methods used in \cite{steorts_2014_hash}, one of which is based upon random projections.
To explore this more broadly, we examine the power of minwise hashing for our sparse representation, while simultaneously utilizing the weighting of various components.

Suppose now $\bx, \by$ are non-negative vectors. For our problem, we are only interested in non-negative vectors because shingle based representations are always non-negative. There is a generalization of Jaccard similarity for real valued vectors in $\mathbb{R}^D$, which unlike minhash is sensitive to the weights of the components, defined as
\begin{align}
 \mathcal{J}_w = \frac{\sum_i \min\{x_i,y_i\}}{ \sum_i \max\{x_i,y_i\}}
 = 1 - \frac{\| \bx - \by \|_1} { \sum_i \max\{x_i,y_i\}},
\end{align}
where $||\cdot||_1$ represents the $\ell_1$ norm.
Consistent weighted sampling~\citep{charikar2002similarity, gollapudi2006exploiting, manasse2010consistent, ioffe2010improved} is used for hashing the weighted Jaccard similarity $J_w$.
In our application to the Syrian database, we find minhash and weighted minhash give similar error rates, which can be seen in \S \ref{sec:app}.


With DOPH  the traditional minwise hashing scheme is linear or constant in the tuning parameters.  
For the weighted version of minhashing, we propose a different way of generating hash values for weighted Jaccard similarity, similar to that of~\cite{charikar2002similarity, gollapudi2006exploiting}. 
As a result, we obtain the fast and practical one pass hashing scheme for generating many different hash values with weights, analogous to DOPH for the unweighted case. Overall, we require only one scan of the record and only one permutation.

Given any two vectors $\bx, \by \in \mathbb{R}^D$ as the shingling representation,
we seek hash functions $h(\cdot)$, such that the collision probability between two hash functions is small. That is,
\begin{equation}Pr(h(x) = h(y)) = \frac{\sum_i {\min\{x_i,y_i\}}}{\sum_i\max\{x_i,y_i\}}.
\end{equation}

Let $\delta$ be a quantity such that all components of any vector $x_i = I_i^x\delta$ for some integer $I_i^x$.\footnote{This assumption is true when dealing with floating point numbers for small enough $\delta$.}  Let  the maximum possible component $x_i$ for any record be $x$  and let $M$ be an integer such that $x_i = M \delta$. Thus, $\delta$ and $M$ always exist for finitely bounded datasets over floating points.

Consider the transformation  $T: \mathbb{R}^D \rightarrow \{0,1\}^{M\times D}$, where for $T(x)$ we expand each component $x_i = I\delta$ to $M$ dimensions and with the first $I$ dimensions have value $1$ and the rest value $0$.

Observe that for vectors $x$ and $y$, $T(x)$ and $T(y)$ are binary vectors and
\begin{align}
&\frac{|T(x) \cap T(y)|}{|T(x) \cup T(y)|} = \frac{\sum_i \min\{I_i^x,I_i^y\}}{ \sum_i \max\{I_i^x,I_i^y\}} \label{eqn:anshu} \\
&= \frac{\sum_i \min\{I_i^x,I_i^y\} \delta}{ \sum_i \max\{I_i^x,I_i^y\} \delta} = \frac{\sum_i {\min\{x_i,y_i\}}}{\sum_i\max\{x_i,y_i\}} \notag
\end{align}
In other words, the usual resemblance (or Jaccard similarity) between the transformed $T(x)$ and $T(y)$ is precisely the weighted Jaccard similarity between $x$ and $y$ that we are interested in.
Thus, we can simply use the DOPH method of~\cite{shrivastava2014densifying,shrivastava2014improved} on $T(x)$ to get an efficient LSH scheme for weighted Jaccard similarity defined by Equation~\ref{eqn:anshu}.  The complexity here is $O(KL+ \sum_i I_i  )$ for generating $k$ hash values, a factor improvement over $O(k\sum_i I_i)$ without the densified scheme.

Often $I_i$ is quite large (when shingling) and $\sum_i I_i$ is large as well.
When $\sum_i I_i$  is large, \cite{gollapudi2006exploiting} give simple and accurate  approximate hashes for weighted Jaccard similarity.  They divide all components $x_i$ by a reasonably big constant so that $x_i \le 1$ for all records $x$. After this normalization, since $x_i \ge 0$, for every $\bx,$ we generate another bag of word $\bx_S$ by sampling each $x_i$ with probability $x_i \le 1$. Then $\bx_S$ is a set (or binary vector) and for any two $\bx$ and $\by$, the resemblance between $\bx_S$ and $\by_S$ sampled in this manner is a very accurate approximation of the weighted Jaccard similarity between $\bx$ and $\by$. After applying the DOPH scheme to the shingled records, we generate $k$ different hash values of each record in time $O(KL+d)$, where $d$ is the number of shingles contained in each record. This is a vast improvement over $O(KL+ \sum_i I_i  )$.  Algorithm~\ref{alg:hashgeneration} summarizes our method for generating $k$ different minhashes needed for blocking.

\begin{algorithm}[h]
\caption{Fast $KL$ hashes}
\label{alg:hashgeneration}
\KwData{record $x$, }
\KwResult{$KL$ hash values for blocking}
 $x_S = \phi$\;
\ForAll{$x_i > 0$}{
 $x_S \cup i$ with probability proportional to $x_i$\;
}
\Return $KL$ densified one permutation hashes (DOPH) of $x_S$
\end{algorithm}

\section{An Application to the Syrian Conflict}
\label{sec:app}
We apply our methods to a database of 296,245 records of identifiable victims that HRDAG has collected from the four aforementioned sources. We consider four sets of features that include  full Arabic name, date of death, high level location (governorate) where the individual died, and sex (M/F). The ultimate blocking method we advocate for (minhashing) is probabilistic and unsupervised.

We analyze the Syrian data using the traditional blocking methods and then by the more advanced hashing-based methods from \S \ref{hashing}. 
We first review how we evaluate our methods using the training data provided. Recall from \S \ref{sec:motivate}, we have hand matched data which we treat as a gold standard against which to compare our blocking approaches.

\paragraph{Evaluation Methods}
We evaluate each of our four hashing methods below using recall and reduction ratio. The recall measures how many of the actual true matching record pairs have been correctly classified as matches. There are four possible classifications. 
First, record pairs can be linked in both the hand-matched training data (which we refer to as `truth') and under the estimated linked data.  We refer to this situation as \emph{true positives} (TP). Second, record pairs can be linked under the truth but \emph{not} linked under the estimate, which are called
\emph{false negatives} (FN).
Third, record pairs can be \emph{not} linked under the truth but linked under the estimate, which are called
\emph{false positives} (FP).
Fourth and finally, record pairs can be \emph{not} linked under the truth and also \emph{not} linked under the estimate, which we refer to as \emph{true negatives} (TN). 
The vast majority of record pairs are classified as true negatives in most practical settings.
Then the true number of links is $\text{TP}+\text{FN}$, while the estimated number of links is $\text{TP}+\text{FP}$.  The usual definitions of false negative rate and false positive rate are
\begin{equation}\notag
\text{FNR}=\frac{\text{FN}}{\text{TP+FN}},\qquad
\text{FPR}=\frac{\text{FP}}{\text{TP+FP}},
\end{equation}
where by convention we take $\text{FPR}=0$ if its numerator and denominator are both zero, i.e., if there are no estimated links.
The recall is defined to be
$$\text{recall} = 1-FNR.$$
The precision is defined to be
$$\text{precision} = 1-FPR.\footnote{Note that the precision for a blocking procedure is not expected to be high since we are only placing similar pair in the same block (not fully running a record linkage procedure or de-duplication procedure, which would try and maximize both the recall and the precision).}$$ 
The reduction ratio (RR) is defined as 
$$RR = 1 - \frac{s_M + s_N}{n_M+n_N},$$ where $n_M$ and $n_N$ are the total of matched and non-matched records and the number of true matched and true non-matched candidate record pairs generated by an indexing technique is denoted with
$s_M + s_N \leq n_M+n_N.$ 
The RR
measure provides information about how many candidate record pairs were generated by an indexing technique compared to all possible record pairs, without assessing the quality of these candidate record pairs. We also evaluate the methods using the precision, where  precision measures the proportion of how many of the classified matches (true positives + false positives) have been correctly classified as true matches (true positives). It thus assesses how precise a classifier is in classifying true matches. 

\subsection{Traditional blocking}
Traditional blocking methods, implementing using the feature set considered here, do not scale to the entire dataset well and more importantly they do not perform well either in terms of recall or the reduction ratio. Based on a subset of 20,000 records from the Syrian database, the recall and reduction ratio is never above 0.30. Thus, we find that we split records that refer to the same individual across different blocks around 70\% of the time.

\subsection{KLSH}
We next apply KLSH, illustrating that application to the Syrian database performs poorly in terms of both recall and reduction ratio. This contrasts 
 empirical studies shown in \cite{steorts_2014_hash}. (The parameters to be set for KLSH are the number of random projections ($p$) and the number of clusters to output ($k$). Through this $k$-means approach to blocking, the mean number of records within a cluster can be fixed. 

\begin{figure}[htbp]]
\begin{center}
\includegraphics[width=0.6\textwidth]{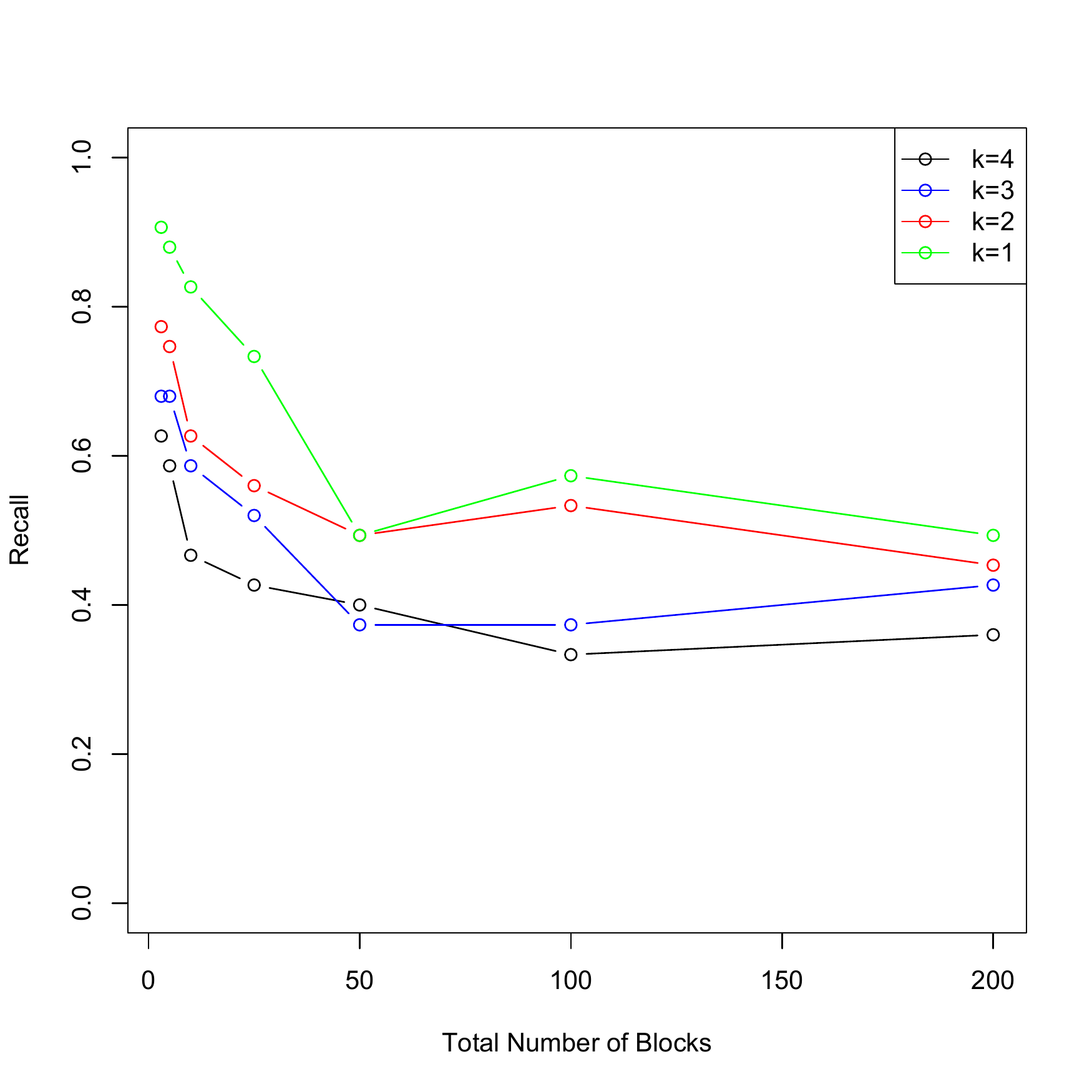}
\caption{KLSH on subset of Syria database (20,000 records) using p=20. 
}
\label{fig:klsh-subset}
\end{center}
\end{figure}

Figure \ref{fig:klsh-subset} displays the results of KLSH clustering on the subset of the Syrian database, where we plot the recall versus the total number of blocks. We set the number of random projections to be $p=20$ and allow the shingles to vary from $k=1,2,3,4.$ This figure shows that a 1-shingle always achieves the highest recall. We notice that using a 1-shingle, a block size of 100, the recall is 0.60, meaning that 40\% of the time we split records referring to the same individual across different blocks. Of course, performing record linkage would not be useful here. 

\begin{figure}[htbp]]
\begin{center}
\includegraphics[width=0.6\textwidth]{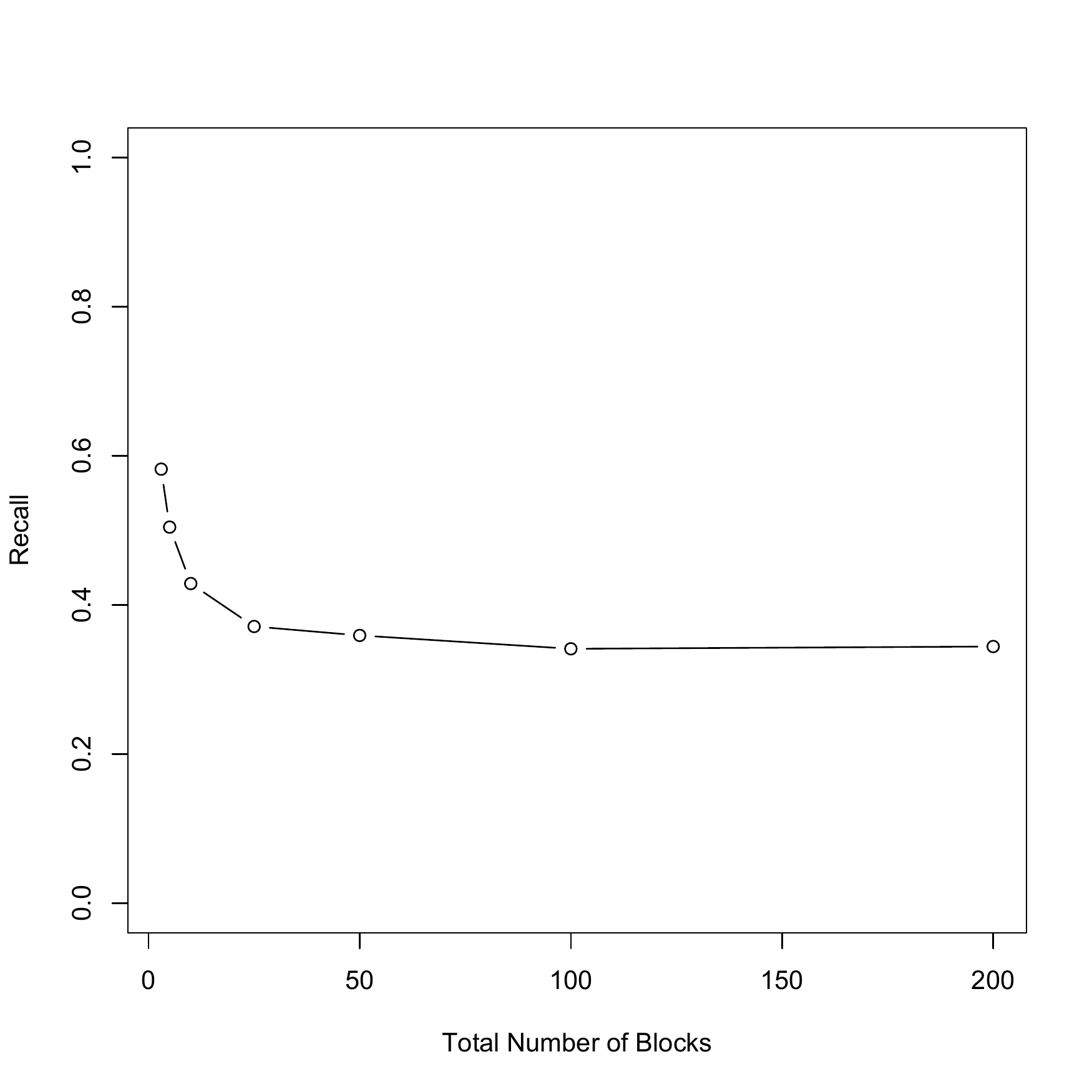}
\caption{KLSH on entire Syria database using p=20.
}
\label{fig:klsh}
\end{center}
\end{figure}

Figure \ref{fig:klsh} displays the results of KLSH clustering on the entire Syrian database, using p=20 and a 1-shingle. We see that when we use all of the data available, the recall decreases as the total number of blocks increases, which happens due to the sparsity and feature-poor data. For the entire Syrian database, a reasonable block size corresponds to a recall of 0.40, meaning that 60\% of the time we split records that refer to the same individual across different blocks. Of course, performing record linkage would not be useful with a recall this low.


\begin{figure}[htbp]
\begin{center}
\includegraphics[width=0.75\textwidth]{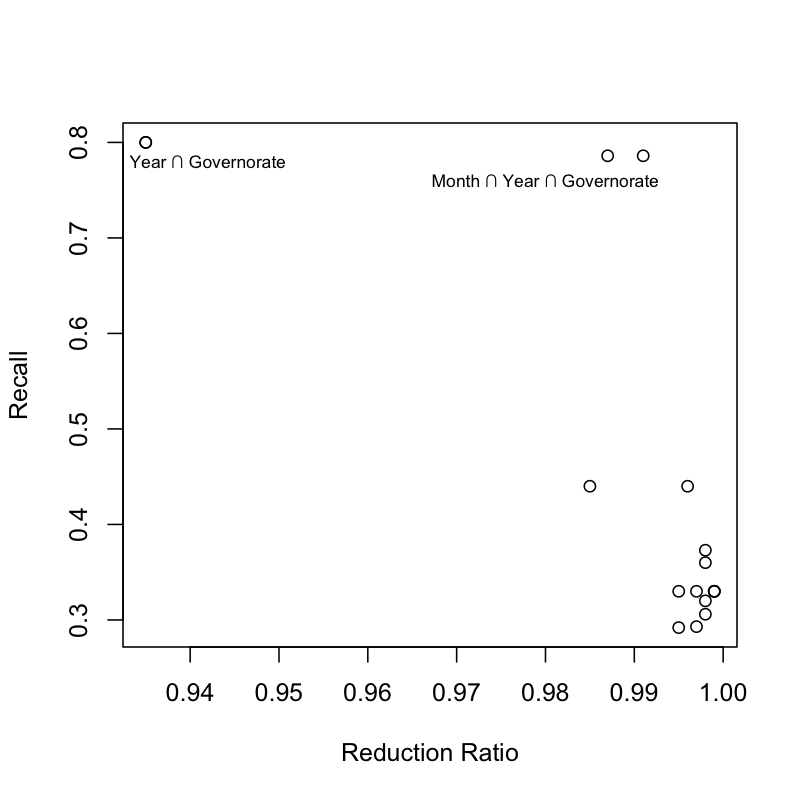}
\caption{Highest performing disjunction of conjunctions sets on entire Syria database, in terms of recall and reduction ratio. The $Year \cap Month \cap Governorate$ blocking is the best result from all of the iterated combinations. We achieve very similar results on the entire Syrian database.}
\label{fig:conj}
\end{center}
\end{figure}

\paragraph{Conjunctions}
Next, we investigate the disjunction of conjunctions method on the subset of the Syrian database. In terms of forming conjunctions, we use date of death, governorate, and full Arabic name. Figure \ref{fig:conj} displays the results of the formed disjunctions on the subset of the Syrian database. We find that $Year \cup Governorate$ and $Month \cup Year \cup Governorate$ produces the highest recall and reduction ratio. 
The highest recall performance is around 0.8, which means that around 80\% of the training pairs are correctly blocked by the conjunction scheme. The results of the conjunction points to the importance of the governorate and date of death fields in classifying records as matches or non-matches.

It is quite easy to calculate the pairs generated by a conjunction blocking scheme. In the conjunction $Year \cup Governorate$, for example, there are 41 blocks created from the total number of records, however these blocks create approximately 1.2 million candidate pairs, which are then compared to the training data. However, there are only 75 training pairs classified as true matches in the Syrian subset of 20,000 records, and so the vast majority of candidate pairs become false positives. Only a few become true positives, and the rest are either false negatives or true negatives. Clearly, the number of candidate pairs only increases in an intractable fashion as the number of records increases. When moving to the entire Syrian database, we find similar results to Figure \ref{fig:conj}.

\subsection{Conjunctions \& Arabic Edit Distance}
We next apply the combined approach of conjunctions and AEDA.
Due to a lack of information regarding data collection methods, the values for the three weighting parameters, $\omega$, $\lambda$, and $\sigma$, are assumed to be of equal importance, and so each is given the value $1/3$. Once a replacement cost value is obtained for each pairing of names within a block, candidate pairs are created by considering the top 10th percentile of records in string similarity (this value can be adjusted to any size, but we found that this value yields the highest recall). 

We perform the AEDA within each block formed by the conjunction rules. The benefit of this approach is that it allows us to take into account the similarities between the Arabic string names to form candidate pairs.  In contrast, in the previous section only exact matches on full names were considered to form simple conjunctions. 

\begin{figure}[htbp]
\begin{center}
\includegraphics[width=0.65\textwidth]{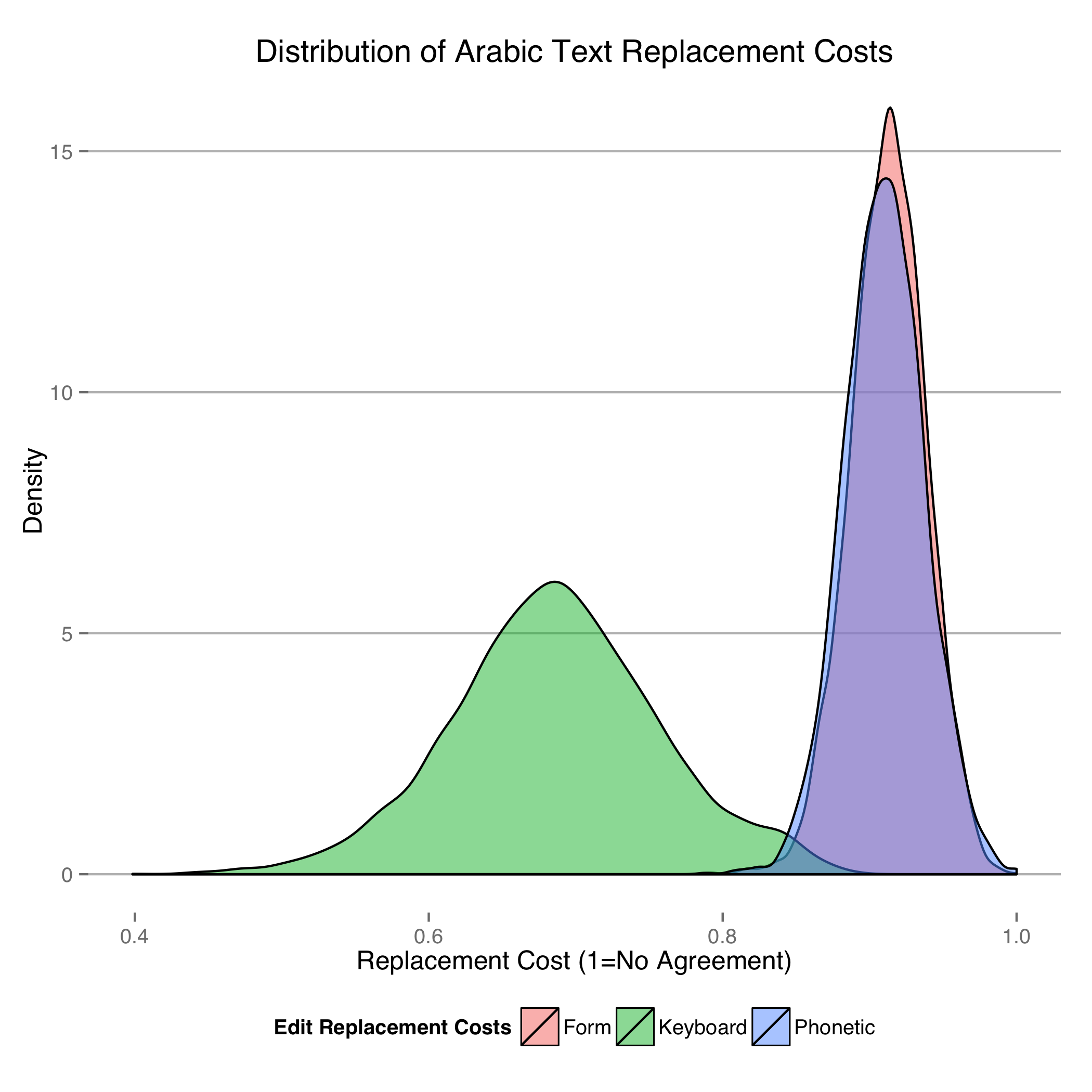}
\caption{Distribution of Arabic text replacement cost measures (perfect matches were removed from the graphic). The replacement cost is $1 - \text{Similarity}$, where $\text{Similarity}$ is how much agreement there is between two given text strings. The keyboard measure results in the greatest differentiation between similar and non-similar Arabic names. The other two measures yield quite similar results, and indicate a low degree of similarity among the tested names. The three distributions are evaluated together by Formula \ref{eqn:fry}, to determine an accurate similarity measure between any two name records. We can also see that the average predicted cost of replacing one string with another is around 0.8, where 1 indicates no agreement at all. This is an indication that the Arabic names present in the databases share very little in common with each other on average.}
\label{fig:arabic}
\end{center}
\end{figure}

First, records are blocked based on a set of features in common. Then, within each block, we compare strings using AEDA. 
The candidate pair generation provided by the AEDA allows for even more reduction in pairwise comparisons, as not all pairs within a conjunction block have to be compared. The results of the AEDA are displayed in Figure \ref{fig:arabic}, which points to the relative importance of each metric in determining similarity.

In terms of recall, combining the AEDA with conjunctions results in equivalent levels, as the same candidate pairs are formed within the cluster as without the differentiation by string. However, the principal advantage of using the AEDA is that the reduction ratio is higher.

\subsection{Minhasing}
\label{sec:minhashing}
\begin{figure}[htbp]
\begin{center}
\includegraphics[width=\textwidth]{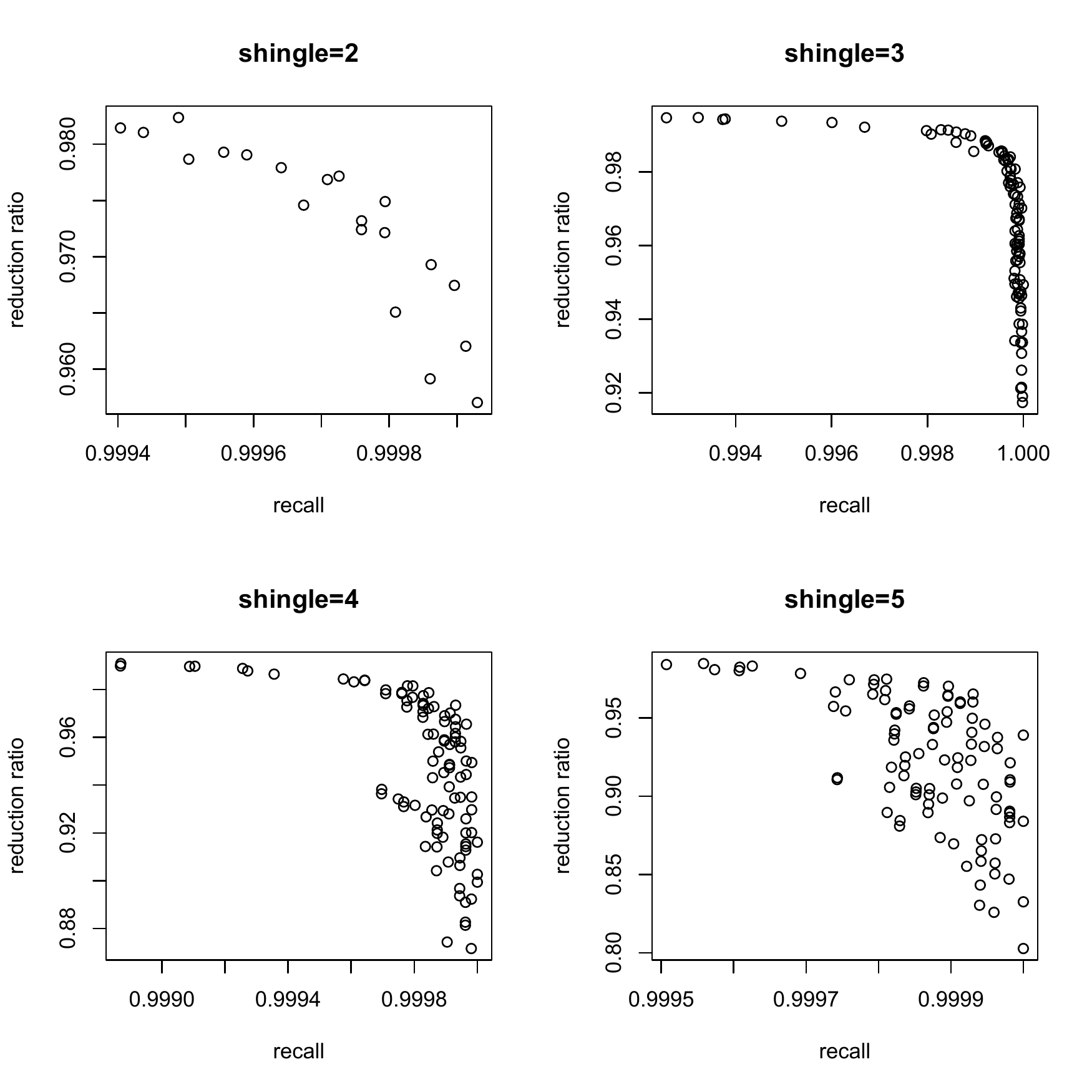}
\caption{For shingles 2--5, we plot the RR versus the recall. Overall, we see the best behavior for a shingle of 3, where the RR and recall can be reached at 0.98 and 1, respectively. We allow L and K to vary on a grid here. L varies from 100--1000 by steps of 100; and K takes values 15, 18, 20, 23, 25, 28, 30, 32, and 35.}
\label{syria-takethatAssad}
\end{center}
\end{figure}

\begin{figure}[htbp]
\begin{center}
\includegraphics[width=\textwidth]{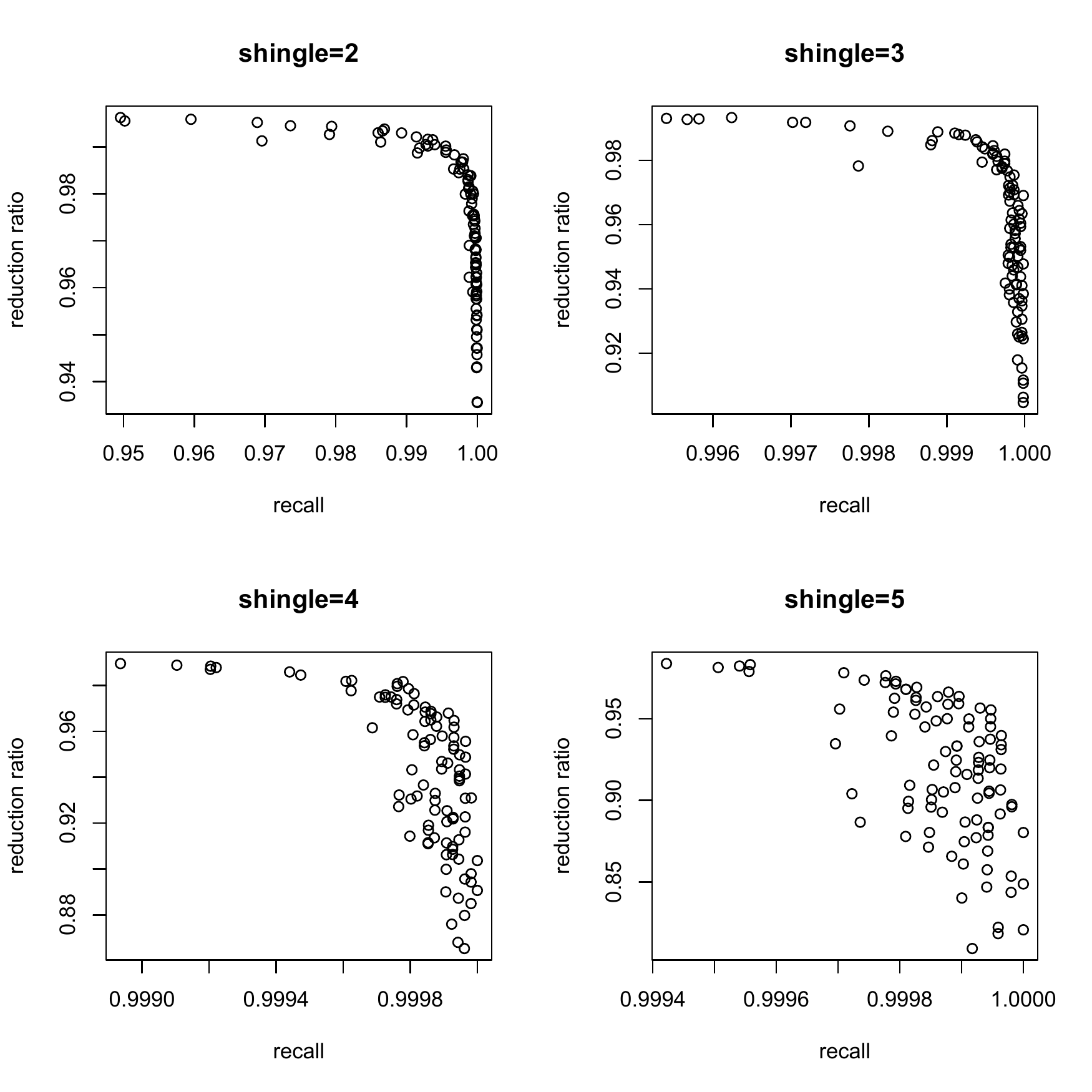}
\caption{For shingles 2--5, we plot the RR versus the recall. Overall, we see the best behavior for a shingle of 2 or 3, where the RR and recall can be reached at 0.98 and 1, respectively. We allow L and K to vary on a grid here. L varies from 100--1000 by steps of 100; and K takes values 15, 18, 20, 23, 25, 28, 30, 32, and 35.}
\label{syria-takethatAssadAgain}
\end{center}
\end{figure}

We apply minhashing using unweighted and weighted DOPH to the full Syrian database using shingles 2---5, where L varies from 100--1000 by steps of 100 and K takes values 15,18,20,23,25,28,30,32,35. We illustrate that regardless of the shingle from 2--5, the recall and RR are close to 1 as illustrated in Figure~\ref{syria-takethatAssad}. Furthermore, using unweighted DOPH, we see that a shingle of three overall is most stable in having a recall and RR close to 0.99 as illustrated in Figure~\ref{syria-takethatAssadAgain}. Using weighted DOPH, we see that a shingle of two or three overall is most stable in having a recall and RR close to 0.99. In terms of computational run time, we note that each individual run takes 10 minutes on the full Syrian dataset. We contrast this with the other blocking runs that on 20,000 records from Syria takes many hours or 1-3 days and return a recall and RR that is unacceptable for record linkage purposes.  While we can achieve a near perfect recall and reduction ratio, the precision is close to 0, meaning that any minhashing method is not sufficient for the purposes of record linkage. Furthermore, the resulting blocks unfortunately overlap and there is no optimal combination of L and K that leads to blocks that do not overlap and have acceptable recall and reduction ratio measures. For the purposes of record linkage, this means that we cannot simply treat each block as ``separate" and run our preferred record linkage procedure in parallel across blocks. Thus, coming up with a reliable estimate of the number of uniquely documented identifiable deaths post-blocking is beyond the scope of the paper and we speak to the potential challenges and solutions to this in \S \ref{sec:disc}

\section{Discussion}
\label{sec:disc}
%

Blocking allows us to reduce the space of all-to-all record comparisons. Specifically, using
locality sensitive hashing, we can effectively block without comprising the recall or RR. Specifically, for minhashing approaches, we have shown that we only split the same record across blocks less than 1\% of the time, where other blocking methods split records across blocks approximately 20 -- 60\% of the time. Furthermore, due to recent computational speedups \citep{shrivastava2014densifying,shrivastava2014improved} we are able to perform one run in 10 minutes, whereas other blocking methods on the same data (or smaller data) take 1--3 days and the accuracy is unacceptable for record linkage purposes. However, in order to reach such high results on both the recall and RR using minhashing, we sacrifice the precision. Hence, hashing based methods are not enough on their own to be used for this application for simultaneous dimension reduction and record linkage. This implies, that based on blocking alone, we cannot achieve a reliable estimate of the observed death count in Syria.

As mentioned in \S \ref{sec:minhashing}, the best blocks we produce are overlapping in the sense that a record can appear in more than one block. When we assess an estimate of the number of uniquely documented identifiable deaths, we seek exact uncertainty quantification from any record linkage procedure, such that the estimate can be as accurate as possible. Currently, exact uncertainty quantification in record linkage is only possible for more than two databases in generative Bayesian methods \citep{steorts_2015_jasa, steorts_2014_aistats, steorts_2015_eb}, where the record linkage is done simultaneously across and within all databases. The immediate challenge is that while we have performed a blocking scheme in 10 minutes, the resulting blocks are overlapping, and hence, any Bayesian method will be computationally very slow on moderately sized data, such as the Syrian dataset. For example, suppose that the largest block contains 500 records. Then the fastest Bayesian record linkage procedure would take about 12--24 hours to run in just one block. Given that there are $M$ blocks, this will take M days to run on one processor (which does not account for running different models, assessing model misspecification, etc.). All in all, such a procedure is computationally intractable without additional computational speed ups.

Perhaps a promising area of exploration is to assume a generative Bayesian record linkage model. We could then combine the Split and Merge method of \cite{smmcmc, steorts_2014_aistats} or the Wormhole method of  \cite{steorts_2015_small_clustering}, with proposals from minhashing to traverse the state space more quickly. This involves testing many record linkage models to see which work best. This not only enables record linkage to be performed, but also the computational complexity of the algorithm could be assessed and multiple systems estimation could also be addressed. Such ideas are very promising in the context of human rights but also in a number of other application areas including official statistics, precision medicine, and genetics.



\subsubsection*{Acknowledgments}
We would like to thank Sam Ventura for comments and help with data cleaning that greatly helped our paper. We would also like to thank Patrick Ball, Jerry Reiter, and Abbas Zaidi for suggestions, comments, and discussions that contributed to the paper. RCS was supported by the John Templeton Foundation and all view expressed by this work are of the authors and not of the foundation.

\clearpage

%
%

\bibliographystyle{imsart-nameyear}
\bibliography{chomp,mybib_merged}

\end{document}